\documentclass[twocolumn]{aa}
\usepackage{natbib}
\bibpunct{(}{)}{;}{a}{}{,}
\usepackage{psfig}
\usepackage{epsfig,afterpage} 
\usepackage{graphicx}
\usepackage{pifont}
\usepackage{amssymb}
\usepackage{array}
\usepackage{amsmath}
\usepackage{rotating}
\usepackage{multirow}
\usepackage{lscape}

\begin{document}

\title  {  Abundances  and  physical parameters for stars in the  open
  clusters  NGC~5822 and  IC~4756.
\thanks{ Based on observations  collected at the European Organisation
  for Astronomical Research in  the Southern Hemisphere, Chile, during
  the observing runs 073.D-0655 and 079.C-0131.
Table  2 is only available   in electronic form at
the CDS via anonymous ftp to cdsarc.u-strasbg.fr (130.79.128.5) or via
http://cdsweb.u-strasbg.fr/cgi-bin/qcat?J/A+A/}}

\subtitle{ }

\author{  G.   Pace \inst{1},  J.   Danziger  \inst{2,3}, G.   Carraro
  \inst{4},   J.   Melendez  \inst{1},   P.   Fran\c   cois  \inst{5},
  F. Matteucci\inst{2,3}, \and N.  C.  Santos \inst{1}}

\offprints{G. Pace, \email gpace@astro.up.pt}

\institute{
            Centro de Astrofisica, Universidade do Porto, 
            Rua das Estrelas, 4150--762 Porto, Portugal    \\  
            \and
            INAF, Osservatorio Astronomico di Trieste, 
            via G.B. Tiepolo 11, 34131 Trieste, Italy\\
            \and
            Department of Astronomy, University of Trieste, 
            via G.B. Tiepolo 11, 34131 Trieste, Italy\\
            \and
            European Southern Observatory, Casilla 19001, Santiago, Chile\\ 
            \and
            Observatoire de  Paris, 64  Avenue de l'Observatoire, 75014
            Paris, France \\}

%\date{Received ...; accepted  ...}

          \abstract{Classical  chemical analyses  may  be affected  by
            systematic  errors  that  would cause  observed  abundance
            differences between dwarfs  and giants. For some elements,
            however,  the  abundance difference  could  be real.}{  We
            address  the issue  by observing  2 solar--type  dwarfs in
            NGC~5822 and 3 in IC~4756, and comparing their composition
            with  that of  3 giants  in either  of  the aforementioned
            clusters.   We   determine  iron  abundance   and  stellar
            parameters  of  the dwarf  stars,  and  the abundances  of
            calcium,  sodium,  nickel,  titanium, aluminium,  chromium,
            silicon, and  oxygen for both the giants  and dwarfs.  For
            the dwarfs, we also  estimate the rotation velocities, and
            and lithium abundances.   We improve the cluster parameter
            estimates  (distance,  age,  and reddening)  by  comparing
            existing  photometry with  new isochrones.}   {We acquired
            UVES  high--resolution,  of  high signal--to--noise  ratio
            (S/N)  spectra.   The   width  of  the  cross  correlation
            profiles  was used  to measure  rotation  velocities.  For
            abundance  determinations, the  standard  equivalent width
            analysis was performed  differentially with respect to the
            Sun.   For lithium  and oxygen,  we derived  abundances by
            comparing synthetic spectra  with observed line features.}
          {We find an iron abundance for dwarf stars equal to solar to
            within  the margins  of  error for  IC~4756, and  slightly
            above   for   NGC~5822   ([Fe/H]=   0.01  and   0.05   dex
            respectively).   The  3 stars  in  NGC  4756 have  lithium
            abundances between Log N(Li)$\approx$ 2.6 and 2.8 dex, the
            two stars in NGC~5822 have Log N(Li)$\approx$ 2.8 and 2.5,
            respectively.  }   {For sodium, silicon,  and titanium, we
            show  that abundances of  giants are  significantly higher
            than those of the dwarfs  of the same cluster (about 0.15,
            0.15,  and 0.35  dex).   Other elements  may also  undergo
            enhancement, but all within  0.1 dex.  Indications of much
            stronger enhancements can  be found using literature data.
            But artifacts  of the  analysis may be  partly responsible
            for this.}


\authorrunning{Pace et al.\ }

\titlerunning{Abundance in solar--type stars in 2 open clusters.}

\maketitle

\section{Introduction}
\label{intro}

Several  projects to  determine open  cluster chemical  abundances and
parameters are being carried out,  such as that of BOCCE \citep{bt06},
in  which  chemical abundances  of  stars in  the  red  clump of  open
clusters are measured,  red clump stars being giants  that are burning
helium  in  their core.   BOCCE,  with  many  singular or  coordinated
studies of stars of different  spectral type and luminosity class, has
significantly enhanced the database  that can be employed to determine
the        Galactic       abundance        gradient       \citep[e.g.,
][]{oldoc,laura09,vale,hyamet}.    These   studies   aim   mainly   to
investigate the evolution with  time of several phenomena occurring in
stars,  such as  the  depletion of  light  elements and  chromospheric
activity,  and global  properties of  the Galactic  disk, such  as the
metallicity    gradient   and   the    age--metallicity   relationship
\citep[e.g.,][]{gmetgr,frmetgr,laura09}.

Targeting open  clusters is justifiable for several  reasons: they are
single  stellar populations, i.e.,  ensembles of  stars with  a common
age, chemical  composition, and distance; they  are distributed across
the entire  Galactic disk; and they  span a wide range  of ages.  Only
very young  clusters tend to cluster  close to the  spiral arms, where
they recently  formed \citep{diaslep}.  High  resolution spectra allow
us  to measure  the  abundance  of light  elements  and in  particular
lithium,  which is  an invaluable  probe of  the mixing  mechanisms in
stellar  envelopes. \cite{Smi}  obtained very  interesting  results by
studying lithium and beryllium in several IC~4651 members at different
evolutionary stages.   Both elements are fragile but  are destroyed at
different temperatures, i.e.,  in different layers, and therefore probe
the stellar interior at different depths.

In addition to the depletion of light elements, we investigate whether
the  chemical composition changes  during evolutionary  transitions in
the stellar  lifetime, for which open cluster  are suited particularly
well.  Sodium  and aluminium have been  found by several  authors to be
overabundant  in  giants  belonging  to  several  open  clusters,  for
instance  IC~4756,   NGC~6939  and  NGC~714   \citep{rv_ic},  NGC~6475
\citep{sandro09},   Collinder   261   \citep{friel03},   Berkeley   17
\citep{friel05},  and Saurer  1 and  Berkeley 29  \citep{giov04}.  For
some  other  clusters, only  sodium  overabundances  were inferred  in
giants:     IC~4651      \citep{luca04},     NGC~7789     and     M~67
\citep{taut1,taut2,linelist}, and NGC~1817 and NGC~2141 \citep{jacob}.
In  some of  these studies,  a direct  comparison between  evolved and
unevolved stars  was possible, and  sodium and aluminium  abundances in
evolved  stars were  found  to be  higher  than in  dwarfs.  In  other
studies, the  overabundances were detected just as  an abundance ratio
[Na--Al/Fe]  greater   than  0.1  dex   in  giants  or   clump  stars.
\cite{linelist}  argue  that  sodium  enhancement in  M~67  should  be
ascribed  to non--LTE effects  affecting sodium  lines in  giants more
than dwarfs, as discussed in \cite{nonlte}. The extreme sensitivity of
sodium to non--LTE effects was also noted by \cite{paola07}. Incorrect
{\textit gf} values may also play a major role for giant stars, since,
unlike dwarf  stars, their differential  analysis with respect  to the
Sun  may not  overcome  this problem.   But  in some  of the  clusters
mentioned above, the  effect may be in part real,  since the amount of
enhancement found  varies significantly from cluster  to cluster, even
when the same  kind of analysis is performed.   In this case, aluminium
and sodium enhancements may be caused by internal processes.

Whether  real or an  artifact of  the analysis,  abundance differences
between  evolved  and  unevolved  stars  have  important  implications
because  they mean  that  spurious differences  may  be found  between
stellar samples, if  care has not been taken  in comparing only dwarfs
with dwarfs, or giants with giants.

In  the present  study, we  analyse 5  solar--type stars  in  the open
clusters IC~4756 and  NGC~5822, and we compare our  results with those
for giants  of the  same clusters.  These  clusters were  studied with
other open clusters by  \cite{giants}.  In this paper, iron abundances
are analysed and we determine the chemical abundances of several other
elements.

The paper is  organised in the following way.  In Sect.  \ref{obs}, we
describe the sample and  the observations.  In Sect.  \ref{chemansec},
we report the chemical abundance measurements with special emphasis on
oxygen and lithium.  An estimate of the projected rotation velocity of
the sample stars is given in Sect.  \ref{vrot}. In Sect.  \ref{cfrpr},
we discuss  the difference between  giants and dwarfs, using  both our
chemical abundance measurements  and previously published results.  In
Sect.   \ref{clpar}, we evaluate  the  cluster  parameters by  fitting
isochrones to  the colour--magnitude diagrams.   Finally, we summarise
the content of the paper in Sect. \ref{conclusions}.

\section{Observations and sample}
\label{obs}

The data  analysed consist  of UVES  spectra of 5  dwarf and  6 giants
members of NGC~5822 and IC~4756. 

The observations  of the  dwarf stars were  designed to  determine the
evolution  in the  chromospheric  activity of  solar--type stars,  and
carried  out  in the  ESO  observing  run  073.D-0655.  The  clusters,
IC~4756 and NGC~5822, were chosen mainly because of their intermediate
age  of about  1 Gyr,  NGC~5822 being  slightly younger  than IC~4756.
Target  stars  were selected  from  all  solar--type, single  members.
Photometric data used in  this selection, including target names, were
taken from  \cite{hs} for IC~4756  and from \cite{tatm}  for NGC~5822.
For the same clusters, UVES spectra of 6 giants were made available to
ourselves.  A  detailed description of  the giant spectra is  given in
\cite{giants} and is not repeated here.

Between 2  and 7 spectra were taken  for each dwarf of  the initial sample
using UVES on the VLT at  a resolution of about R=~100\,000 in the red
arm,  which covers  the range  from  4\,800 to  6\,800 {\AA}.   Radial
velocity   measurements  were   used   to  strengthen   single--member
selection.  They  were obtained by  cross--correlating stellar spectra
with that of the Sun using the IRAF command {\textit fxcor}.

We detected  the following double  stars in IC~4756:  HER~40, HER~150,
HER~183, HER~189, and HER~294. The stars HER~150, HER~189, and HER~294
are double--lined  and the other  binaries were detected because  of a
radial velocity variation larger  than 1 km$\cdot$sec$^{-1}$.  We were
left  with only  3 single  members  for this  cluster, namely  HER~97,
HER~165, and HER~240,  which each had 2 observations  on two different
nights: their radial velocity differences of, respectively, 0.06, 0.20
and 0.40  km$\cdot$sec$^{-1}$ are attributable  to measurement errors.
Their radial  velocities are -24.9,  -25.7, -24.9 km$\cdot$sec$^{-1}$,
which confirm  their membership when assuming the  mean cluster radial
velocity of -25.2 km$\cdot$sec$^{-1}$ \citep{rv_ic,mm08}.  The cluster
is affected  by variable extinction \citep{schm78};  therefore the use
of photometric  information to select  the targets \citep{hs}  may not
have prevented a large proportion  of double stars from being included
in the original sample.

After adding the spectra, the  resulting S/N per pixel for these stars
was  about 50  in the  blue arm  and 100  in the  red arm.  The HER~97
spectra are  of slightly  poorer quality than  those of the  other two
stars,  which probably explains  the larger  difference between  the 2
radial velocity measurements.

In NGC~5822,  we observed TATM~11014 seven times,  and TATM~11003 four
times in a  total of 6 different nights.   Their radial velocities for
the  individual observations have  a standard  deviation of  about 0.5
km$\cdot$sec$^{-1}$  for  both  stars,   which  we  attribute  to  the
measurement errors.  The S/N of the single spectra can be, in fact, as
low as 20.

The radial  velocity of  TATM~11003 is -29.1  km$\cdot$sec$^{-1}$, and
that  of TATM~11014  is  -27.9 km$\cdot$sec$^{-1}$.   The mean  radial
velocities  of  the  20  confirmed  members  of  NGC~5822  studied  by
\cite{mm90}   have  a   distribution   whose  mean   value  is   -29.3
km$\cdot$sec$^{-1}$,   with  a   standard  deviation   of   about  0.8
km$\cdot$sec$^{-1}$,  thus  confirming the  membership  of our  target
stars.  After coadding all the spectra of each star, we achieved a S/N
of 50  for the blue arm of  TATM~11014 and slightly more  than 100 for
the red  arm for the same  star, and about 60  and 30 for  the red and
blue arm of TATM~11003, respectively. We used some of the spectra used
in \cite{paper1}, including the  solar spectrum from the UVES archive,
to  calibrate the  rotation velocity  (Sect.  \ref{vrot})  and measure
lithium abundances (Sect. \ref{liab}).

\section{Chemical abundances}
\label{chemansec}

To measure  the stellar parameters of dwarf  stars, i.e., temperature,
gravity, microturbulent velocity, and the abundances of iron, calcium,
sodium,  nickel,  titanium,   aluminium,  chromium,  and  silicon,  we
employed the same procedure  and error analysis used in \cite{papoc1},
to which we  refer the reader for a  detailed description.  Suffice it
to say that we performed a standard EW analysis, i.e., we measured EWs
and inferred abundances for each  individual line (in the case of iron
in   two  ionization   states)  by   means  of   OSMARCS   LTE  models
\citep{osmarcs}, differentially line by  line with respect to the Sun,
and that the  adopted atmospheric parameters were chosen  among a grid
of trial values.  The only  difference between the present analysis of
dwarf  stars  and  that  of  \cite{papoc1}, is  that  the  temperature
estimates about  which the grids of parameters  were constructed, were
obtained from  the B-V colours published in  \cite{hs} and \cite{tatm}
through  the  calibrations in  \cite{bvtcal}  and \cite{bvtcal2}.   To
deredden  the   colours,  we  used  the  E(B-V)   values  computed  in
Sect. \ref{clpar}.

\begin{table*}
\begin{center}
\begin{tabular}{c c c c c c c c c c c c c} 
&\multicolumn{3}{|c|}{T$_{eff}$}&\multicolumn{3}{|c}{$\log$G}\\
Star&%
\multicolumn{1}{|c}{S08}&\multicolumn{1}{c}{HM}&\multicolumn{1}{c}{R06}&
\multicolumn{1}{|c}{S08}&\multicolumn{1}{c}{HM}&\multicolumn{1}{c}{R06}\\
\hline
\hline
\\
IC  4756 No38 &5225$\pm$26&5151$\pm$73&5226&3.16$\pm$0.22&3.16$\pm$0.17&3.09\\
IC  4756 No42 &5240$\pm$26&5217$\pm$89&5231&3.14$\pm$0.22&3.21$\pm$0.17&3.18\\
IC  4756 No125&5207$\pm$26&5146$\pm$82&5257&3.06$\pm$0.26&3.11$\pm$0.18&3.13\\
NGC 5822 No102&5253$\pm$28&5170$\pm$42&5260&3.17$\pm$0.39&3.20$\pm$0.13&3.28\\
NGC 5822 No224&5214$\pm$28&5237$\pm$65&5277&3.14$\pm$0.20&3.37$\pm$0.12&3.29\\
NGC 5822 No438&5208$\pm$25&5148$\pm$62&5208&3.16$\pm$0.19&3.21$\pm$0.11&3.19\\
\\
&    \multicolumn{3}{|c}{$\xi$}&\multicolumn{3}{|c}{[Fe/H]}\\
&%
\multicolumn{1}{|c}{S08}&\multicolumn{1}{c}{HM}&\multicolumn{1}{c}{R06}&
\multicolumn{1}{|c}{S08}&\multicolumn{1}{c}{HM}&\multicolumn{1}{c}{R06}\\
\hline
\hline
\\
IC  4756 No38 &1.40$\pm$0.02&1.22$\pm$0.11&1.35&0.05$\pm$0.08&0.08$\pm$0.11&0.09\\
IC  4756 No42 &1.39$\pm$0.02&1.15$\pm$0.13&1.26&0.01$\pm$0.08&0.10$\pm$0.11&0.12\\
IC  4756 No125&1.47$\pm$0.02&1.31$\pm$0.13&1.23&0.02$\pm$0.08&0.07$\pm$0.11&0.10\\
NGC 5822 No102&1.44$\pm$0.03&1.18$\pm$0.07&1.28&0.00$\pm$0.08&0.05$\pm$0.06&0.04\\
NGC 5822 No224&1.41$\pm$0.03&1.15$\pm$0.08&1.29&0.06$\pm$0.08&0.22$\pm$0.08&0.21\\
NGC 5822 No438&1.39$\pm$0.02&1.13$\pm$0.08&1.19&0.06$\pm$0.08&0.18$\pm$0.08&0.14\\
\\

\end{tabular}
\caption{Comparison of  abundance and  parameter measurements for giants.}
\end{center}
\label{cfrgiants} 
\end{table*}

As  for  giant stars,  stellar  parameters  and  iron abundances  were
determined   in  \cite{giants}   using   the  line   list  of   either
\citeauthor{s08}  (\citeyear{s08},  S08)  or that  of  \citeauthor{HM}
(\citeyear{HM}, HM).   Effective temperature and  gravity measurements
in the two  sets agree to within the margins of  error, while the iron
abundances obtained using HM are between 1 and 2 $\sigma$ higher. 
  We  determined the  stellar  parameters for  giants  using the  same
  procedure and  line list  as for  dwarfs, and we  show in  Table~1 a
  comparison  between  our results,  labelled  ``R06'',  and those  of
  \cite{giants}, labelled  either ``S08''  or ``HM'' according  to the
  line list  adopted.   Our  measurements match the  results obtained
using HM, and we used them to  proceed, in the same way as for dwarfs,
with the measurement of  calcium, sodium, nickel, titanium, aluminium,
chromium,  silicon  and  oxygen  abundances.   In  \cite{giants},  the
preliminary results  of the dwarf iron abundances  presented here were
compared to the giant abundances.  We  do not have new elements to add
to that discussion, and we do not repeat it here.

\addtocounter{table}{1}

The  equivalent   width  measurements  and   the  relative  individual
abundances are available at CDS in electronic form (Table 2).  Some of
them are made automatically with  the program ARES \cite{ARES}.  For a
couple  of   stars,  we   verified  that  automatic   and  interactive
measurements are  in no  poorer agreement than  two different  sets of
measurements performed  using the interactive  method.  The parameters
found  are indicated in  Table \ref{starpar}.  In Cols.   2 and  3, we
indicate the  photometry from \cite{tatm} and  \cite{hs}.  To deredden
B-V, we used the colour excesses obtained in Sect. \ref{clpar}. In the
Cols.  4 and  5, we  indicate the  temperature obtained  from  the B-V
colour   by   means  of   the   calibrations   of  \cite{bvtcal}   and
\cite{bvtcal2},  while in  Col. 6  we indicate  those obtained  in the
spectroscopic  analysis.   The   difference  between  photometric  and
spectroscopic   temperature  is   extremely  large   in   all  dwarfs,
spectroscopic temperature being in all cases higher, as was mostly the
case  in  \cite{papoc1}.   If  we  consider  photometric  temperatures
resulting  from the  calibration in  \cite{bvtcal},  these differences
range from  about 150 K to 600  K.  Expressed in terms  of B-V colour,
the mismatch amounts to about 0.13 mag for both stars in NGC~5822, and
range from 0.05 to 0.17 mag for the stars in IC~4756.  The uncertainty
in the colour excess computed in Sect. \ref{clpar} amounts to 0.05 and
0.10  mag  for NGC~5822  and  IC~4756,  respectively.   These and  the
uncertainties   in  the  (B-V)--versus--temperature   calibration  are
unlikely to be the main cause of the discrepancy between spectroscopic
and photometric temperatures, since,  for instance, the calibration in
\cite{bvtcal2} has a  standard deviation of about 50  K.  For IC~4756,
differential  reddening  \citep{schm78}  and  the  larger  photometric
errors may account  for the mismatch. The spread  in the main sequence
is of  the order of  0.2 mag.  In  both clusters, there is  probably a
colour  offset  in  the  adopted  photometry, possibly  caused  by  an
inaccurate  transformation  from  the  instrumental  to  the  standard
magnitude system.   A zeropoint difference  of about 0.1 mag  or more,
which  is necessary to  explain the  mismatch between  photometric and
spectroscopic  temperature, is  not rare  for  photographic photometry
when compared to modern CCD studies, and this illustrates the need for
new CCD  photometry for  both clusters.  In  this study,  we therefore
rely  on the  spectroscopic determinations  of temperature,  which are
definitely more trustworthy.

\begin{figure}
\begin{center}
\includegraphics[width=9cm]{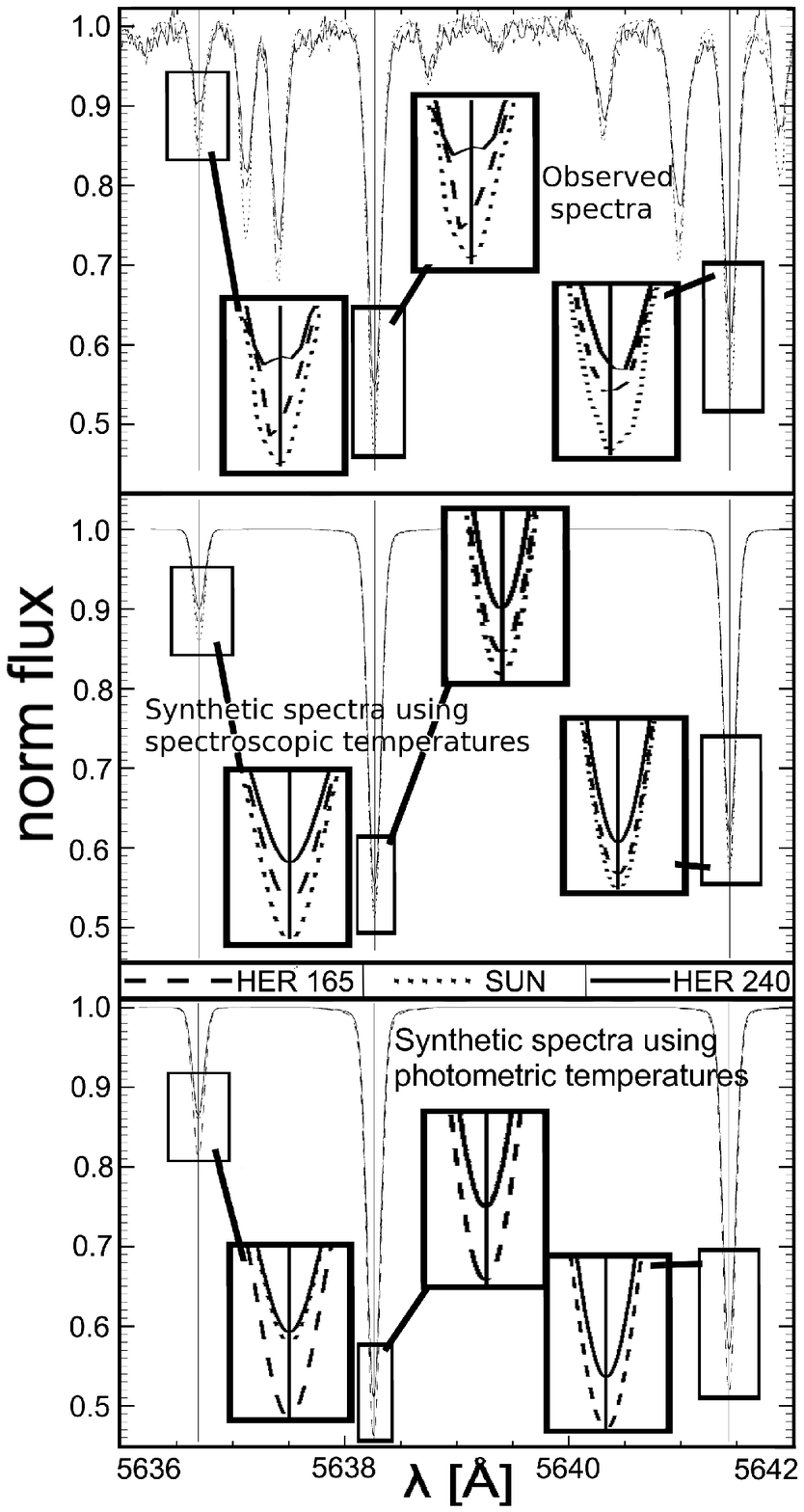}\\
\end{center}
\caption{Comparison of the spectra of HER~240 and HER~165 with that of
  the Sun.}
\label{fig1}
\end{figure}

To  corroborate the  last statement,  we show  in Fig.   \ref{fig1} a
comparison between 3 iron lines in the spectra of the Sun and those of
HER~165 and HER~240, predicted to be close to solar temperature on the
basis of the  photometric calibrations of, respectively, \cite{bvtcal}
and \cite{bvtcal2}.  Our  spectroscopic analysis indicates, instead, a
higher temperature. In the top panel of Fig.  \ref{fig1} the observed
spectra are plotted, in the  central panel we show synthetic models of
the  iron  lines obtained  by  using,  for  HER~240 and  HER~165,  the
parameters  found in  the spectroscopic  analysis, and  in  the bottom
panel  we  show  the  same  spectral synthesis  but  with  photometric
temperature instead of the  adopted one.  The synthetic profiles based
on spectroscopic temperatures resemble  much more closely the observed
ones.  In particular, we note that, assuming a photometric temperature
for HER 240, its spectrum is  expected to match almost exactly that of
the Sun,  which is not the  case.  The region indicated  was chosen to
contain several Fe {\sc I} lines in a narrow range.

\begin{table*}
\begin{center}
\begin{tabular}{c c c c c c c c}
Star&\multicolumn{1}{|c}{V}  & \multicolumn{1}{c|}{ B-V } &\multicolumn{1}{|c}{ T$_{Soder.}$}&T$_{Casag.}$%
& \multicolumn{1}{c|}{T$_{spec}$}& \multicolumn{1}{|c|}{log G}& \multicolumn{1}{|c}{$\xi$} \\
    &\multicolumn{1}{|c}{} & \multicolumn{1}{c|}{}  &\multicolumn{3}{|c|}{}&&  \multicolumn{1}{|c}{}\\
    &\multicolumn{1}{|c}{} & \multicolumn{1}{c|}{}  & &&  \multicolumn{3}{c}{[$\log (g \ cm \ sec^2$)]}\\
    &\multicolumn{2}{|c|}{mag}  & \multicolumn{3}{|c|}{[K]}&  &\multicolumn{1}{|c}{[km$\cdot$sec$^{-1}$]}\\
\hline
\\
\multicolumn{8}{l}{\bf IC 4756}\\
HER  97   & 13.38 & 0.85& 5512 & 5418 & 6118    &4.46 & 1.29 \\
HER 165   & 13.50 & 0.81& 5658 & 5562 & 6070    &4.45 & 1.24 \\ 
HER 240   & 13.48 & 0.76& 5848 & 5753 & 6007    &4.54 & 1.21 \\ 
\\
\multicolumn{8}{l}{\bf NGC 5822}\\
TATM 11003& 14.662& 0.769& 5625& 5546 & 6160    &4.74 & 1.05 \\
TATM 11014& 14.448& 0.732& 5764& 5685 & 6273    &4.74 & 1.26 \\
\hline
\end{tabular}
\end{center}
\caption{Stellar  parameters.}
\label{starpar}
\end{table*}

\begin{table*}
\begin{center}
\begin{tabular}{c c c c c c c c c c c c c}

Star &\multicolumn{3}{|c}{Fe {\sc I}}&\multicolumn{3}{|c}{Na {\sc I}}&\multicolumn{3}{|c}{Al {\sc I}}&\multicolumn{3}{|c|}{Si {\sc I}}\\
or &\multicolumn{3}{|c}{}&\multicolumn{3}{|c}{}&\multicolumn{3}{|c}{}&\multicolumn{3}{|c|}{}\\
Cluster name&\multicolumn{1}{|c}{[X/H]}&N&$\sigma$&\multicolumn{1}{|c}{[X/H]}&N&$\sigma$&\multicolumn{1}{|c}{[X/H]}&N&$\sigma$&\multicolumn{1}{|c}{[X/H]}&N&\multicolumn{1}{c|}{$\sigma$}\\
%\hline
\\
\multicolumn{13}{c}{dwarfs}\\
    HER 165&    0.05$\pm$    0.09& 62&    0.04&   -0.10$\pm$    0.06&  3&    0.03&   -0.01$\pm$    0.05&  1&    0.00&    0.01$\pm$    0.02&  8&    0.04\\
    HER 240&   -0.02$\pm$    0.09& 62&    0.05&   -0.17$\pm$    0.06&  3&    0.06&   -0.20$\pm$    0.05&  2&    0.04&   -0.06$\pm$    0.02&  8&    0.03\\
    HER  97&    0.00$\pm$    0.09& 61&    0.06&   -0.18$\pm$    0.06&  3&    0.08&   -0.06$\pm$    0.05&  1&    0.00&   -0.07$\pm$    0.03&  8&    0.03\\
\\
\multicolumn{13}{c}{mean dwarf}\\
  IC 4756  & 0.01&\multicolumn{2}{c}{}&-0.15&\multicolumn{2}{c}{}&-0.09&\multicolumn{2}{c}{}&-0.04&\multicolumn{2}{c}{}\\
\\
\multicolumn{13}{c}{giants}\\
    No  38 &    0.08$\pm$    0.11& 16&    0.08&    0.19$\pm$0.04&  3&    0.04 &     -0.04$\pm$0.03&  2&    0.03&    0.10$\pm$0.02&  9&    0.06\\  
    No  42 &    0.10$\pm$    0.11& 15&    0.08&    0.23$\pm$0.03&  3&    0.05 &     -0.03$\pm$0.03&  2&    0.05&    0.11$\pm$0.02&  9&    0.06\\
    No 125 &    0.07$\pm$    0.11& 15&    0.08&    0.16$\pm$0.04&  3&    0.04 &     -0.03$\pm$0.03&  2&    0.04&    0.09$\pm$0.02&  9&    0.05\\
\\
\multicolumn{13}{c}{mean giant}\\
  IC 4756  & 0.08&\multicolumn{2}{c}{}& 0.19&\multicolumn{2}{c}{}&-0.03&\multicolumn{2}{c}{}& 0.10&\multicolumn{2}{c}{}\\
\\
\multicolumn{13}{c}{dwarfs}\\
     TATM 11014&    0.07$\pm$    0.09& 49&    0.05&   -0.11$\pm$    0.06&  3&    0.05&   -0.05$\pm$    0.05&  1&    0.00&    0.04$\pm$    0.03&  6&    0.03\\
     TATM 11003&    0.02$\pm$    0.09& 57&    0.07&   -0.19$\pm$    0.06&  2&    0.03&    --               & --&  --    &   -0.01$\pm$    0.03&  7&    0.05\\
\\
\multicolumn{13}{c}{mean dwarf}\\
NGC 5822&0.05&\multicolumn{2}{c}{}&-0.15&\multicolumn{2}{c}{}&-0.05&\multicolumn{2}{c}{}& 0.01&\multicolumn{2}{c}{}\\
\\
\multicolumn{13}{c}{giants}\\
       No 102  &    0.05$\pm$    0.06& 13&    0.04&    0.15$\pm$0.03&  3&    0.09&   -0.04$\pm$0.02&  2&  0.09  &    0.03$\pm$0.02&  9&    0.09\\
       No 224  &    0.22$\pm$    0.08& 14&    0.06&    0.26$\pm$0.03&  3&    0.05&    0.05$\pm$0.02&  2&  0.06  &    0.18$\pm$0.02&  9&    0.07\\
       No 438  &    0.18$\pm$    0.08& 16&    0.06&    0.23$\pm$0.03&  3&    0.05&    0.05$\pm$0.02&  2&  0.05  &    0.20$\pm$0.02&  9&    0.12\\
\\
\multicolumn{13}{c}{mean giant}\\
NGC 5822&0.15&\multicolumn{2}{c}{}& 0.21&\multicolumn{2}{c}{}& 0.02&\multicolumn{2}{c}{}& 0.14&\multicolumn{2}{c}{}\\
\multicolumn{13}{c}{}\\
%\hline
%\multicolumn{13}{c}{}\\
     &\multicolumn{3}{|c}{Ca {\sc I}}&\multicolumn{3}{|c}{Ti {\sc I}}&\multicolumn{3}{|c}{Cr {\sc I}}&\multicolumn{3}{|c|}{Ni {\sc I}}\\
% Star&\multicolumn{3}{|c}{Ca {\sc I}}&\multicolumn{3}{|c}{Ti {\sc I}}&\multicolumn{3}{|c}{Cr {\sc I}}&\multicolumn{3}{|c|}{Ni {\sc I}}\\
%or &\multicolumn{3}{|c}{}&\multicolumn{3}{|c}{}&\multicolumn{3}{|c}{}&\multicolumn{3}{|c|}{}\\
%Cluster name&\multicolumn{1}{|c}{[X/H]}&N&$\sigma$&\multicolumn{1}{|c}{[X/H]}&N&$\sigma$&\multicolumn{1}{|c}{[X/H]}&N&$\sigma$&\multicolumn{1}{|c}{[X/H]}&N&\multicolumn{1}{c|}{$\sigma$}\\
%\hline
\\
\multicolumn{13}{c}{dwarfs}\\
       HER 165&    0.10$\pm$    0.08&  9&    0.04&    0.02$\pm$    0.10& 11&    0.06&    0.05$\pm$    0.11&  6&    0.05&   -0.03$\pm$    0.06& 21&    0.05\\
       HER 240&   -0.01$\pm$    0.08& 11&    0.03&   -0.05$\pm$    0.10& 10&    0.06&    0.00$\pm$    0.11&  6&    0.05&   -0.09$\pm$    0.06& 22&    0.05\\
        HER 97&    0.06$\pm$    0.08& 11&    0.06&   -0.08$\pm$    0.10& 11&    0.12&   -0.04$\pm$    0.11&  5&    0.04&   -0.07$\pm$    0.07& 20&    0.07\\
\\
\multicolumn{13}{c}{mean dwarf}\\
IC 4756  &0.05&\multicolumn{2}{c}{} &-0.04&\multicolumn{2}{c}{}& 0.00&\multicolumn{2}{c}{}&-0.06&\multicolumn{2}{c}{}\\
\\
\multicolumn{13}{c}{giants}\\
    No  38 &    0.06$\pm$0.05& 9 &    0.04&    0.11$\pm$0.07&  9&    0.08&    0.08$\pm$0.07&  6&    0.03&    0.03$\pm$0.03& 23&    0.07\\  
    No  42 &    0.09$\pm$0.05& 9 &    0.05&    0.15$\pm$0.07&  9&    0.09&    0.14$\pm$0.07&  6&    0.03&    0.05$\pm$0.03& 23&    0.07\\
    No 125 &    0.03$\pm$0.05& 9 &    0.05&    0.08$\pm$0.07&  9&    0.08&    0.05$\pm$0.07&  6&    0.04&    0.05$\pm$0.03& 23&    0.08\\
\\
\multicolumn{13}{c}{mean giant}\\
  IC 4756  & 0.06&\multicolumn{2}{c}{}& 0.11&\multicolumn{2}{c}{}& 0.08&\multicolumn{2}{c}{}& 0.04&\multicolumn{2}{c}{}\\
\\
\multicolumn{13}{c}{dwarfs}\\
    TATM 11014&    0.09$\pm$    0.08&  9&    0.05&    0.01$\pm$    0.09&  6&    0.07&    0.11$\pm$    0.10&  6&    0.04&   -0.01$\pm$    0.06& 21&    0.06\\
    TATM 11003&    0.06$\pm$    0.08&  9&    0.04&   -0.02$\pm$    0.10&  9&    0.04&    0.04$\pm$    0.11&  6&    0.10&   -0.06$\pm$    0.06& 19&    0.07\\
\\
\multicolumn{13}{c}{mean dwarf}\\
 NGC 5822& 0.08&\multicolumn{2}{c}{}& 0.00&\multicolumn{2}{c}{}& 0.07&\multicolumn{2}{c}{}&-0.03&\multicolumn{2}{c}{}\\
\\
\multicolumn{13}{c}{giants}\\
       No 102  &    0.02$\pm$0.03&  9&    0.06&    0.07$\pm$0.05&  9&   0.08 &    0.09$\pm$0.05&  6&  0.05  &   -0.11$\pm$0.02& 23&    0.09\\
       No 224  &    0.16$\pm$0.03&  9&    0.07&    0.28$\pm$0.05&  9&   0.08 &    0.24$\pm$0.05&  6&  0.03  &    0.17$\pm$0.02& 23&    0.10\\
       No 438  &    0.14$\pm$0.03&  9&    0.05&    0.19$\pm$0.05&  9&   0.09 &    0.21$\pm$0.05&  6&  0.03  &    0.14$\pm$0.02& 23&    0.09\\
\\
\multicolumn{13}{c}{mean giant}\\
NGC 5822&0.11&\multicolumn{2}{c}{}& 0.18&\multicolumn{2}{c}{}& 0.18&\multicolumn{2}{c}{}& 0.07&\multicolumn{2}{c}{}\\
%\hline
\end{tabular}
\end{center}
\caption{Results of the chemical  analysis. }
\label{chemtab}
\end{table*}

 The results  of the  chemical analysis  are summarised  in Table
  \ref{chemtab}.  Errors  in  the  [X/H]  values  are  those  computed
  considering the uncertainty in the  parameters and in the EWs, while
  the  $\sigma$  columns  indicate   the  standard  deviation  of  the
  measurements  from individual  lines. When  only  2 or  3 lines  are
  present,  the  standard  deviation   is  replaced  by  half  of  the
  difference  between  the  largest   and  the  smallest  value.  Iron
  abundances   in   giant  stars   and   relative   errors  are   from
  \cite{giants}. 

\subsection{Oxygen abundances}

To determine oxygen  abundances, the only line available  to us is the
forbidden O  {\sc I}  line at  6300.30 {\AA}, which  is very  weak and
often  contaminated by telluric  lines.  We  analysed this  line using
MOOG  \citep[][version 2002]{MOOG}  and  Kurucz models  \citep{kur93}.
The line is blended with that of  Ni {\sc I} at 6300.336 {\AA}.  As in
\cite{linelist}  and \cite{papoc1},  we  assumed a  log~$gf$ value  of
-2.11 for the  oxygen line and and -9.717 for that  of the nickel.  To
measure abundances  in dwarfs, we  selected the spectra for  which the
telluric lines  did not overlap  with the oxygen profile.   Even after
summing these  spectra, the  S/N ratio at  the wavelength  of interest
remained too low to reliably measure the EW of the the blended feature
or to compare  the spectra with a synthetic model.   We could only set
an  upper limit.   Comparing  the  stellar feature  with  that of  the
UVES--archive solar  spectrum, we  noted that, in  all the  cases, the
solar feature  was well suited as  an upper limit to  the stellar one.
We employed  the driver  {\textit synthe} of  MOOG, assuming  as fixed
input  the  stellar  parameters  and  nickel  abundances  obtained  as
described above, to compile the  models of our stars; we then searched
for the oxygen abundances for which the synthetic spectrum matched the
solar one  most closely,  thus finding an  upper limit to  the stellar
oxygen abundance.  For the Sun itself, the match between synthetic and
observed     profile     was     achieved     by     assuming     that
log~(O/H)+12=8.83$\pm$0.03.   This value was  subtracted from  all the
stellar  values previously obtained,  to determine  an upper  limit to
[O/H].  For  all the 3  stars in IC~4756,  we claim that  [O/H] $\leq$
0.15 dex.   For the spectrum of  HER~165, which is similar  to that of
the Sun,  we estimate that  the oxygen abundance  of this star  is not
much  lower than  the aforementioned  upper limit,  but, owing  to the
scantiness of  the data on which  it depends, this  conclusion must be
interpreted with caution.  For NGC~5822  dwarfs, we could find only an
upper limit of 0.3 dex.

Despite the spectra for giants  at our disposal being of high quality,
we could not  obtain precise measurements because the  oxygen line was
blended with  the telluric feature  at 6299 {\AA}, and  no calibration
target was available to us.  The centre of the telluric line coincided
exactly with that of the oxygen line in the IC~4756 spectra, producing
a  single Gaussian  feature.  In  the NGC~5822  spectra, the  two line
centers were  separated sufficiently  to create a  slightly asymmetric
feature, but in no case was it possible to achieve reliable results by
the comparing  the observed spectra with the  synthetic ones. However,
the telluric line  at 6302 {\AA} was isolated, and  its EW ranged from
15 to 25 m{\AA} in all  the spectra.  The telluric line at 6299 {\AA},
which is blended with the oxygen  line, is a few percent stronger.  We
thus inferred  that oxygen  abundances in giants  of both  IC~4756 and
NGC~5822 must be between -0.1 and 0.15 dex.  Unfortunately, the direct
comparison of  giants and dwarfs in  the same cluster on  the basis of
these approximations, does not add relevant information.

\subsection{Lithium }
\label{liab}

Lithium abundances  were obtained by analysing the  lithium doublet at
6707.8 {\AA}. Adopting a procedure similar to that used for oxygen, we
employed  the driver  {\textit synthe}  of  MOOG, the  line list  from
\cite{lilist}, and Kurucz models  obtained with the stellar parameters
measured  above,  to  produce  synthetic  spectra  in  the  wavelength
interval encompassing  the lithium line profile, and  searched for the
lithium abundance for which the  closest match to the observed feature
is achieved.  We were able to obtain reliable measurements because the
profiles are quite strong in our stars. The typical errors, across the
lithium  abundance  range for  which  the  match  was acceptable,  are
between  0.05 and  0.15 dex.   If we  were to  adopt  the conservative
estimate  of   110  K  for  the   error  in  the   temperature  as  in
\cite{papoc1},  this  would lead  to  an  uncertainty  in the  lithium
abundance of  0.1 dex,  which would imply  a typical maximum  error of
about  0.15  dex.  The  other  parameters  hardly  affect the  lithium
abundance determination.   Using the same procedure,  we also compared
the  synthetic  spectrum  of  the  Sun with  the  UVES--archive  solar
spectrum,  finding that  we  obtained the  closest  match by  assuming
A(Li)= 1  rather than  the canonical A(Li)=1.1  where A(Li) is,  as is
customary, log(Li/H)+12.

Since we believe that a more precise temperature scale may improve our
understanding of  lithium depletion (Sect.  \ref{liab})  and errors in
the  colour excess  may introduce  a cluster  to cluster  bias  in the
temperature determinations (cf.  Sect.  \ref{chemansec}), we measured
the lithium  abundances of  stars in the  sample of  \cite{papoc1} for
which data  in the literature were based  on photometric temperatures,
and whose  parent clusters are significantly  reddened and photometric
temperatures, as  a consequence, were  uncertain.  The results  of our
lithium abundance measurements are shown in Table \ref{litab}, in dex,
along  with temperature,  in Kelvin,  and gravity,  in unit  of $\log$
(g$\cdot$cm$\cdot$sec$^{-2}$).  Since, using  the same method, the Sun
is found to  have A(Li)=1.0, the data are  probably more comparable to
literature sources when a positive offset of 0.1 dex is added.

\begin{figure}
\begin{center}
\includegraphics[width=8cm]{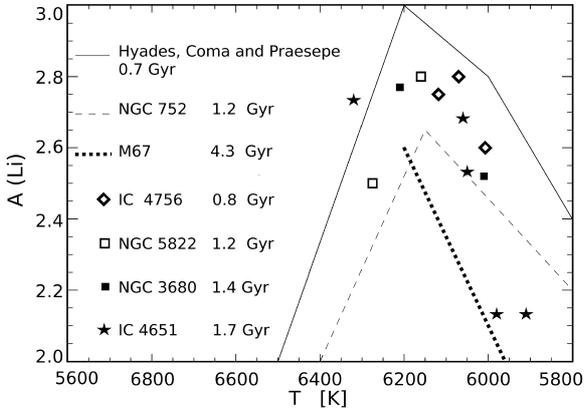}
\caption{Comparison  of  lithium  abundances  in different open clusters.}
\label{fig3} 
\end{center}
\end{figure}

In  Fig.   \ref{fig3},  we  also  show a  temperature  versus  lithium
abundance diagram,  in which  we compare the  12 data points  from the
present  analysis, shown  in  Table \ref{litab},  with published  open
cluster  data  at 3  different  ages. Data  for  M~67  are taken  from
\cite{lcbiaz},   the   remainder  being   from   the  compilation   in
\cite{xiong}.   For clarity, literature data were represented with
  curves  instead of  datapoints, drawn  by eye  in the  region around
  which the datapoints would cluster. Hyades, Coma, and Praesepe data
are  depicted in  a single  curve, another  represents NGC~752,  and a
third  M~67.   The ages  indicated  are  taken  from \cite{sal}.   The
temperatures for the Hyades age  clusters and NGC~752 are derived from
photometry.   For the  former, the  uncertainty in  the  colour excess
should not play a major role, since they are nearby clusters.

The additional 12 points introduced  in Fig. \ref{fig3} clearly do not
account for all unsolved problems in mixing and lithium depletion
  in solar--type stars.  Nevertheless, the steeper decline of lithium
abundance in IC~4651 suggests that a study of a larger sample of stars
in this cluster would be warranted.

\begin{table}
\begin{center}
\begin{tabular}{c c c c}
Star&A(Li)& T$_{eff}$&$\log$G\\
\hline
\hline
\multicolumn{4}{c}{}\\
\multicolumn{4}{c}{Present sample}\\
\multicolumn{4}{c}{}\\
\hline
\hline
\multicolumn{4}{c}{}\\
\multicolumn{4}{c}{IC 4756 \ \  ([Fe/H]=0.01 dex)}\\
\multicolumn{4}{c}{}\\
    HER  97    & 2.75& 6118&4.46\\  
    HER 165    & 2.8 & 6070&4.45\\    
    HER 240    & 2.6 & 6007&4.54\\
\hline
\multicolumn{3}{c}{}\\
\multicolumn{3}{c}{NGC 5822 \ \  ([Fe/H]=0.05 dex)}\\
\multicolumn{3}{c}{}  \\
 TATM 11003    &  2.8 & 6160&4.74\\    
 TATM 11014    &  2.5 & 6273&4.74\\     
\hline
\hline
\multicolumn{4}{c}{}\\
\multicolumn{4}{c}{Sample of \cite{papoc1}}\\
\multicolumn{4}{c}{}\\
\hline
\hline
\multicolumn{4}{c}{}\\
\multicolumn{4}{c}{IC 4651 \ \  ([Fe/H]=0.12 dex)}\\
\multicolumn{4}{c}{}\\
 AHTC 1109     & 2.7& 6060 &4.55\\    
 AHTC 2207     & 2.5& 6050 &4.36\\  
 AHTC 4220     & 2.1& 5910 &4.57\\
 AHTC 4226     & 2.1& 5980 &4.44\\
 Eggen 45      & 2.7& 6320 &4.43\\
 \hline
\multicolumn{4}{c}{}\\
\multicolumn{4}{c}{NGC 3680 \ \  ([Fe/H]=-0.04 dex)}\\
\multicolumn{4}{c}{}\\
 Eggen 70      & 2.8& 6210 &4.47\\     
 AMC 1009      & 2.5& 6010 &4.50\\    
\hline
\end{tabular}
\caption{Lithium   abundance   measurements.}
\end{center}
\label{litab} 
\end{table}

\section{Rotation velocity}
\label{vrot}

In  \cite{paper1} we obtained, using  available  UVES  spectra,   the
cross--correlation profile  with a  suitable template (courtesy  of C.
Melo) and measured its FWHM for  a set of stars with published $v\cdot
\sin i$ taken with the same configuration as for those of the targets;
we found  the values of $A$ and $B$ that  minimise the $\chi^2$ in
the relationship
$$ v\cdot \sin i = A \cdot \sqrt{FWHM^2-B^2}.$$
We then  used that  relationship with the  computed values of  $A$ and
$B$, to  measure target--star $v\cdot \sin  i$ from the  FWHM of their
cross--correlation profile.  For the  present analysis we repeated the
calibration    with   the    same   stars,      obtaining  the
cross--correlation profiles  of calibration and target  stars with the
IRAF command {\textit  fxcor} using the upper red  arm of UVES spectra
(from 5800 to  6800 {\AA}) and, as a template,  the UVES archive solar
spectrum.   This  new  calibration  made our  measurements  easier  to
reproduce.  Calibration data --  namely star names, FWHM, $v \sin
  i$, and deviations from the fit -- are shown in Table \ref{caltab}.
We  found the  following values  for  the parameters  in the  equation
above:  $A=$ 0.82  [km$\cdot$sec$^{-1}$], $B=$  5.93 pixels,  giving a
root  mean  square   of  the  data  points  around   the  fit  of  1.5
km$\cdot$sec$^{-1}$.   We  adopt this  value  as  an  estimate of  the
uncertainty  in the  $v \sin  i$  measurement due  to the  calibration
$\Delta_{cal}$

\begin{table}
\begin{center}
\begin{tabular}{c c c c}
  Star           &  FWHM   & $v \sin i$&  dev.     \\ 
  & [pixels]& [km$\cdot$sec$^{-1}$]  & [km$\cdot$sec$^{-1}$] \\
  \hline
  \\
  Sun                 &  5.46 &  2.0 &  -2.0 \\ 
\\
  \multicolumn{4}{l}{stars in IC 4651: }\\
  AT 1228    & 17.74 & 15.7 &  -1.9 \\
  Eggen 15   & 11.76 & 10.0 &  -1.6 \\
  Eggen 34   & 32.19 & 25.2 &   0.9 \\
  Eggen 45   &  7.56 &  4.2 &  -0.3 \\
  Eggen 79   & 25.6  & 21.8 &  -1.3 \\
  Eggen 99   & 36.28 & 28.1 &   1.4 \\
\\ 
 \multicolumn{4}{l}{star in NGC 3680: }\\
  Eggen 60  &  7.10 &  1.9 &   1.3 \\    
  \hline
\end{tabular}
\end{center}
\caption{Data used to calibrate the $v \sin i$.}
\label{caltab}
\end{table}

The  results for  projected  rotation velocities  are  given in  Table
\ref{vsinitab}. We also  computed the error in FWHM  from the standard
deviation  of  the  FWHM  measurements  performed  on  the  individual
spectra, and  when only two spectra  were available for  a given star,
half of  the difference between  the two measurements were  used.  The
error  in  FWHM  measurement  was   found  to  be  much  smaller  than
$\Delta_{cal}$. However,  the results  provide only a  rough estimate.
The stars  of our sample appear  to rotate slightly  more rapidly than
the Sun, for which  $v \sin i$=2~km$\cdot$sec$^{-1}$, and similarly to
the solar stars in IC~4651 and NGC~3680.  These slow rotators in young
clusters are  not surprising, since \cite{paper1}  measured low values
of $v \sin  i$ also in Hyades and Praesepe  stars.  The data presented
in Table \ref{vsinitab}  alone, is not sufficient to  allow us to draw
conclusions  about the  evolution of  angular momentum  in  solar type
stars.

\begin{table}
\begin{center}
\begin{tabular}{c c c}
  Star& FWHM&$v \sin i$\\
  &[pixels]&[km$\cdot$sec$^{-1}$]\\
  \hline
  \multicolumn{3}{c}{}\\
  \multicolumn{3}{c}{IC 4756}\\
  \multicolumn{3}{c}{}\\
  HER 165    &   6.9   &    3\\    
  HER  97    &   8.1   &    5\\  
  HER 240    &   7.0   &    3\\
  \hline
  \multicolumn{3}{c}{}\\
  \multicolumn{3}{c}{NGC 5822}\\
  \multicolumn{3}{c}{}\\
  TATM 11014    &   8.9   &    5\\     
  TATM 11003    &   7.1   &    3\\    
  \hline
\end{tabular}
\end{center}
\caption{$v  \sin  i$ estimations  of  our  target  stars.  }
\label{vsinitab}
\end{table}

\section{Comparison of our chemical abundances with previous results}
\label{cfrpr}

\begin{table*}
\begin{center}
\begin{tabular}{c c c c c c c}
  
Cluster   &   source    &[Fe/H]             &[Na/H]               &[Al/H]              &[Si/H]                &  Ref \\
\hline
\\
  IC 4756 & 3 dwarfs    & 0.01 $\sigma$=0.04& -0.15 $\sigma$= 0.04& -0.09 $\sigma$=0.05& -0.04 $\sigma$=0.04  &  1   \\
  IC 4756 & 1 dwarf     & 0.03  --          &                     &                    &                      &  4   \\
\\
  IC 4756 & 3 giants    & 0.08 $\sigma$=0.02&  0.19 $\sigma$= 0.04& -0.03 $\sigma$=0.05&  0.10 $\sigma$=0.01  &  1,2 \\
  IC 4756 & 7 giants    & 0.0  $\pm$0.1   * &                     &                    &                      &  3   \\
  IC 4756 & 4 giants    &-0.05 $\pm$0.04    &                     &                    &                      &  4   \\
  IC 4756 & 1 giant     &                   &                     &0.20                &0.16$\pm$0.25         &  4   \\
  IC 4756 & 6 giants    &-0.15 $\sigma$=0.04&0.73 $\sigma$= 0.06 *&0.44 $\sigma$=0.08 *&  0.19 $\sigma$=0.06  &  5   \\
\\
\\
 NGC 5822 & 2 dwarfs    & 0.05 $\sigma$=0.03& -0.15 $\sigma$= 0.04& -0.05 --           &  0.01 $\sigma$=0.02  &  1   \\
\\
 NGC 5822 & 3 giants    & 0.15 $\sigma$=0.08&  0.21 $\sigma$= 0.04&  0.02 $\sigma$=0.01&  0.14 $\sigma$=0.08  &  1,2 \\
 NGC 5822 & 3 giants    & 0.12 $\sigma$=0.1 &                     &                    &                      &  4   \\
 NGC 5822 & 1 giant     &                   & 0.28$\pm$0.07       &0.12$\pm$0.12       &0.25$\pm$0.25         &  4   \\
\\
          &             &[Ca/H]             &[Ti/H]               &[Cr/H]              &[Ni/H]                &      \\
\hline
\\
  IC 4756 & 3 dwarfs    & 0.05 $\sigma$=0.05& -0.04 $\sigma$= 0.05&  0.00 $\sigma$=0.05& -0.06 $\sigma$=0.03  &  1   \\
\\
  IC 4756 & 3 giants    & 0.06 $\sigma$=0.03&  0.11 $\sigma$= 0.03&  0.08 $\sigma$=0.03&  0.04 $\sigma$=0.01  &  1,2 \\
  IC 4756 & 1 giant     &-0.06$\pm$0.29     & -0.28$\pm$0.29      &                    &0.04$\pm$0.16         &  4   \\
  IC 4756 & 6 giants    &-0.08 $\sigma$=0.08&                     &                    & -0.07 $\sigma$=0.05  &  5   \\
\\
\\
 NGC 5822 & 2 dwarfs    & 0.08 $\sigma$=0.02&  0.00 $\sigma$= 0.02&  0.07 $\sigma$=0.04& -0.03 $\sigma$=0.03  &  1   \\
\\
 NGC 5822 & 3 giants    & 0.11 $\sigma$=0.07&  0.18 $\sigma$= 0.11&  0.18 $\sigma$=0.08&  0.07 $\sigma$=0.14  &  1,2 \\
 NGC 5822 & 1 giant     &-0.05$\pm$0.23     &  0.20$\pm$0.24      &0.26$\pm$0.29       &0.25$\pm$0.26         &  4   \\
%\hline
\end{tabular}
\end{center}
\caption{Compilation  of abundance  determinations  in the  literature
  from  spectroscopic data.
  References:
  (1) Present work;
  (2) \cite{giants};
  (3) \cite{groy};
  (4) \cite{luck};
  (5) \cite{rv_ic}
}
\label{cfrtab}
\end{table*}

In Table \ref{cfrtab}, we compiled abundance measurements available in
the literature,  by means of  spectroscopic studies. We  indicated the
dispersion in the measurements,  using either their standard deviation
or,  when  only  3 or  2  measurements  were  available, half  of  the
difference  between the  highest and  lowest value.   The measurements
given  by  \cite{groy}  (flagged  with  an  asterisk)  have  only  two
significant  digits,  and  for 5  out  of  7  stars resulted  to  have
[Fe/H]=0.0  dex.    The  value  indicated   in  this  table   is  more
representative  of her  results than  the mean  of the  7 measurements
(which would  be 0.04 dex).   The measurements by  \cite{rv_ic} marked
with an asterisk  refer to the EW analysis  result. Spectral synthesis
indicate an enhancement 0.06 dex  higher for sodium and 0.11 dex lower
for aluminium.

We note that the iron  abundance measurement of giants in IC~4756 made
by \cite{rv_ic} are substantially  lower than any other quoted result,
including that of the present analysis and of \cite{giants}.  Jacobson
et   al.    employed   Hydra/WIYN   spectra   at   a   resolution   of
R$\approx$15\,000, whose S/N range from 75 to 150 per pixel, and whose
spectral coverage is 300 {\AA}, enough to include more than 20 Fe {\sc
  i} lines and 3 or 4 Fe {\sc ii} lines.  Furthermore, they analysed 6
giants, and their measurements show little spread.  \cite{ARES} showed
that, for a  spectrum of a dwarf  with S/N of about 100  per pixel, at
resolutions  lower  than  R$\approx$30\,000,  EWs  are  systematically
underestimated.  For R=15\,000, this  effect is about 10\%, leading to
abundance  errors that  might  account for  the  mismatch between  our
results and those of Jacobson.   Giants are likely to be more strongly
affected,  because  of  their  higher  line  crowding.   However,  for
abundance ratios [X/Fe], these effects, which affect both sides of the
ratio, should compensate each other.

Data shown in  Table \ref{cfrtab} indicate that there  is a difference
in the chemical composition between giants and dwarfs.  In particular,
sodium  abundance  is  significantly  enhanced in  giants.   The  huge
discrepancy between different enhancements found, e.g., in the present
work and  \cite{rv_ic}, is mainly due  to the use of  a different line
list.  These authors claim that were they to use the same line list we
employed \citep{linelist},  they would find  [Na/H]=0.2, which matches
our result for giants.

\section{Revision of cluster  fundamental parameters}
\label{clpar}

\begin{figure}
\begin{center}
\includegraphics[width=8cm]{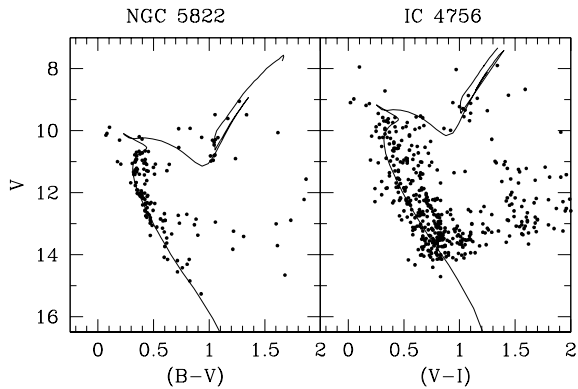}
\caption{CMD and fit of  isochrones for NGC~5822 and IC~4756.} 
\label{fig2} 
\end{center}
\end{figure}

The new iron  abundance estimates that we provide  in this paper offer
us the  possibility of revising  the cluster fundamental  parameters -
namely  distance,  reddening  and  age   -  on  a  more  solid  basis.
Therefore, we  collated the available photometry  from the literature,
and  created colour  magnitude diagrams  (CMD).  We  adopted \cite{hs}
photographic photometry for IC 4756 and \cite{tatm} CCD photometry for
NGC~5822.  The two CMDs are shown  in the right and left panel of Fig.
\ref{fig2}, respectively.  We  transformed iron abundances values from
the logarithmic  scale relative  to the Sun  ([Fe/H]) into  the linear
scale  (Z)   using  the  relation   [Fe/H]~=~log~$({Z}/{0.019})$  from
\cite{glc99}.  We then  generated isochrones for that value  of Z.  We
obtained Z =  0.019 for IC 4756 and  Z = 0.021 for NGC  5822.  In Fig.
\ref{fig2}, we show  the best fit {\it by eye}  that we achieved after
several trials.  The  fit to the CMD of NGC 5822  (left panel) is good
providing  the set  of parameters:  E(B-V)  = 0.1$\pm$  0.05, (m-M)  =
9.9$\pm$0.1,  and age  =  1.0$\pm$0.1  Gyr. This,  in  turn, yields  a
heliocentric distance of 830 pc.  The CMD of IC~4765 has a much larger
spread  and a  secondary main  sequence (MS),  presumably  produced by
binary stars,  and which seems to  be as populated as  the single star
MS.   For this  cluster, our  parameters are  E(B-V)  = 0.15$\pm$0.10,
(m-M) = 8.6$\pm$0.1, and age  = 0.8$\pm$0.2 Gyr. This, in turn, yields
a heliocentric distance of 430 pc.

As  noted  in Sect.   \ref{chemansec},  the  huge differences  between
photometric  and spectroscopic temperatures  imply that  a zero--point
error exist in the photometry of roughly 0.13 mag for NGC 5822, and of
the same order of magnitude for IC~4756 but more difficult to evaluate
due to  the differential reddening.   This can explain  the difference
between the extinctions obtained in this section and those obtained in
the extinction maps of \cite{hak} as described in \cite{jorge06}.  The
aforementioned zero--point error affects the present evaluation of the
cluster parameters by an amount  that should be added to random errors
(the errors given above include  it).  The age evaluation, however, is
not significantly affected by this.

In both cases, we highlight  the need for improved photometry to reach
fainter magnitudes  along the MS,  and comparison fields to  deal with
contamination.   These are  of  primary importance  for  at least  two
reasons.  More precise  age estimates will firstly allow  us to either
confirm or revise the  conclusions reached in \cite{letter} concerning
the evolution of chromospheric activity.  Secondly, we will be able to
say  whether the  cause  of the  discrepancy  between photometric  and
spectroscopic temperatures resides completely in the photometry, as we
suspect,  or   whether  a   systematic  error  in   our  spectroscopic
temperature determinations is present and should be corrected for.

\section{Conclusions.}
\label{conclusions}

We have  analysed high  resolution spectra of  5 solar--type  stars in
IC~4756 and NGC~5822 to obtain their parameters, chemical composition,
including  lithium  abundances,   and  estimates  of  their  projected
rotation velocities.   While our iron  abundances, in most  cases, are
consistent  within the  errors,  with those  of  previous studies  for
giants  of the same  clusters, sodium, aluminium and  silicon, present
very high enhancements in giants in IC~4756.  This finding agrees with
published  results   of  several   other  open  clusters   (see  Sect.
\ref{intro}).  Regardless of  whether these abundance enhancements are
real  or due to  systematic errors,  they have  important implications
\citep{desilva}; for example in the studies of chemical evolution
  of the Galactic  bulge, when comparing the bulge  with the disk, the
  same type of stars in both populations should be used.  In the first
  studies of  this kind employing  high--resolution spectroscopy, this
  was   not  yet  possible   for  a   sufficient  number   of  targets
  \citep{fulbright,lecbulge,manbulge}.    But   works  are   presently
  available   that    compare   bulge   giants    with   disk   giants
  \citep[e.g.,][]{jorgeblg,alan,ryde}   or  bulge  dwarfs   with  disk
  dwarfs,    taking    advantage    of   the    microlensing    effect
  \citep[e.g.,][]{bensby}. The rotation  velocity of our sample stars
is  slightly larger  than that  of the  Sun, roughly  between 3  and 5
km$\cdot$sec$^{-1}$, a  smaller range than observed  in other clusters
of comparable age.

Cluster  parameters are  computed using  published photometry  and the
iron abundance obtained in the spectroscopic analysis. We found an age
of 1~Gyr for NGC~5822 and 0.8~Gyr for IC~4756. However, for IC~4756 in
particular,   new    photometry   would   considerably    reduce   the
uncertainties.  This is  of crucial importance owing to  the impact of
the ages of these clusters on the characterisation of the evolution of
chromospheric activity.

We  compared  lithium abundances  with  published  cluster  data at  3
different ages,  and with measurements  made on stars  in intermediate
age  clusters from  our sample  of \cite{papoc1}.   Most of  the stars
studied  by us  have lithium  abundance levels  lying between  that of
NGC~752 and  that of the  Hyades--age clusters.  Even though  based on
only 4  datapoints, we  see evidence that  a steep decline  in lithium
abundances occurs below 6\,000 K in IC~4651.

Accurate determinations  of temperature and lithium abundances
in more member stars of such  clusters as IC~4651 would clearly be 
invaluable in
understanding the reason for differences among clusters.

\begin{acknowledgements}

  The quality  of the paper improved  due to the  valuable comments of
  the referee,  L. Pasquini.  G.  P.  acknowledges  L.  Casagrande, L.
  Girardi, M.  Zoccali and  S.  Recchi for  useful discussions, F. Pires
  for his help on the layout of Fig. 1,  and S.
  Sousa  for his kind  support on  ARES, his  code for  the authomatic
  measurement of  the equivalent widths.  The  data analysis performed
  here benefited from the  impressive simplicity of use and efficiency
  of ARES.  Data was collected at ESO, VLT.  This publication made use
  of  data  products  from   the  WEBDA  database,  created  by  J.-C-
  Mermilliod and  now operated at  the institute for Astronomy  of the
  University  of Vienna.   The  SIMBAD astronomical  database and  the
  NASA's   Astrophysics  Data  System   Abstract  Service   were  also
  extensively used.   G.  P.  acknowledges the support  of the Funda\c
  c\~{a}o  para a Ci\^{e}ncia  e a  Tecnologia (Portugal)  through the
  grant SFRH/BPD/39254/2007 and the project PTDC/CTE-AST/65971/2006.

\end{acknowledgements}

%####

\bibliographystyle{aa}

\end{document}